\documentclass[fleqn,usenatbib]{mnras}

\usepackage{newtxtext,newtxmath}

\usepackage[T1]{fontenc}

\DeclareRobustCommand{\VAN}[3]{#2}
\let\VANthebibliography\thebibliography
\def\thebibliography{\DeclareRobustCommand{\VAN}[3]{##3}\VANthebibliography}

\usepackage{graphicx}	
\usepackage{amsmath}	

\newcommand{\hi}{\mbox{H\,{\sc i}}}

\newcommand{\mgi}{\mbox{Mg\,{\sc i}}}

\newcommand{\civ}{\mbox{C\,{\sc iv}}}
\def\lya{\ensuremath{{\rm Ly}\alpha}}

\def\h2{H$_2$}

\def\kms{km\,s$^{-1}$}
\def\cms{cm$^{-2}$}

\def\zabs{$z_{\rm abs}$}
\def\zem{$z_{\rm em}$}
\def\nhi{$N$($\hi$)}
\def\ts{$T_{\rm s}$}

\def\v90{$v_{\rm 90}$}

\def\J23{J2358$-$1020}

\title[Variable \hi\ 21-cm absorption]{Emergence of a new \hi\ 21-cm absorption component at $z \sim$1.1726 towards the $\gamma-$ray blazar PKS~2355-106
}

\author[Srianand et al.]{Raghunathan Srianand $^{1}$\thanks{E-mail:anand@iucaa.in}, Neeraj Gupta$^1$, Patrick Petitjean$^2$, Emmanuel Momjian$^3$, Sergei A. Balashev$^{4,5}$ \newauthor{Fran\c{c}oise Combes$^6$, Hsiao-Wen Chen $^7$,  Jens-Kristian Krogager$^8$, Pasquier Noterdaeme$^{2,9}$, Hadi Rahmani$^{10}$,}\newauthor{Andrew J. Baker$^{11,12}$,   Kimberly L. Emig$^{13}$\thanks{Jansky Fellow of the National Radio Astronomy Observatory}, Gyula I. G. J\'ozsa$^{14,15}$, Hans-Rainer Kloeckner$^{14}$,  \& Kavilan Moodley$^{16,17}$
}
\\
\\
$^{1}$ IUCAA, Postbag 4, Ganeshkind, Pune 411007, India\\
$^{2}$ Institut d’Astrophysique de Paris, Sorbonne Université and CNRS, 98bis boulevard Arago, F-75014 Paris, France\\
$^3$ National Radio Astronomy Observatory, P.O. Box O, Socorro, NM 87801, USA\\
$^4$ Ioffe Institute, Politekhnicheskaya 26, 194021 Saint Petersburg, Russia\\
$^5$ HSE University, Saint Petersburg, Russia \\
$^6$ Observatoire de Paris, Collège de France, PSL University, Sorbonne University, CNRS, LERMA, Paris, France\\
$^7$ Department of Astronomy \& Astrophysics, The University of Chicago, 5640 South Ellis Avenue, Chicago, IL 60637, USA \\
$^8$ Université Lyon1, ENS de Lyon, CNRS, Centre de Recherche Astrophysique de Lyon UMR5574, F-69230 Saint-Genis-Laval, France\\
$^{9}$ Franco-Chilean Laboratory for Astronomy, IRL 3386, CNRS and U. de Chile, Casilla 36-D, Santiago, Chile\\
$^{10}$ School of Astronomy, Institute for Research in Fundamental Sciences (IPM), P.O. Box 19395-5531, Tehran, Iran\\
$^{11}$ Department of Physics and Astronomy, Rutgers, the State University of New Jersey, 136 Frelinghuysen Road, Piscataway, NJ
08854-8019, USA\\
$^{12}$ Department of Physics and Astronomy, University of the Western Cape, Robert Sobukwe Road, Bellville 7535, South Africa\\
$^{13}$National Radio Astronomy Observatory, 520 Edgemont Road, Charlottesville, VA 22903, USA
\\
$^{14}$ Max-Planck-Institut f\"ur Radioastronomie, Auf dem H\"ugel 69, D-53121 Bonn, Germany\\
$^{15}$ Department of Physics and Electronics, Rhodes University, P.O. Box 94, Makhanda, 6140, South Africa \\
$^{16}$ Astrophysics Research Centre, University of KwaZulu-Natal, Westville Campus, Durban 4041, South Africa\\
$^{17}$ School of Mathematics, Statistics \& Computer Science, University of KwaZulu-Natal, Westville Campus, Durban 4041, South Africa\\
}
\date{Accepted XXX. Received YYY; in original form ZZZ}

\pubyear{2021}

\begin{document}
\label{firstpage}
\pagerange{\pageref{firstpage}--\pageref{lastpage}}
\maketitle

\begin{abstract}
We report the emergence of a new \hi\ 21-cm absorption at \zabs = 1.172635 in the damped \lya\ absorber (DLA) towards the $\gamma$-ray blazar { PKS~2355-106} (\zem $\sim$1.639) using science verification observations (June 2020) from the MeerKAT Absorption Line Survey (MALS).  
Since 2006, this DLA is known to show a narrow \hi\ 21-cm absorption at \zabs = 1.173019 coinciding with a distinct metal absorption line component.
We do not detect significant \hi\ 21-cm optical depth variations from this known \hi\ component.
A high resolution optical spectrum (August 2010) shows a distinct Mg~{\sc i} absorption at the redshift of the new \hi\ 21-cm absorber. However, this component is not evident in the profiles of singly ionized species. We measure the metallicity ([Zn/H] = $-$(0.77$\pm$0.11) and [Si/H]= $-$(0.96$\pm$0.11)) and depletion ([Fe/Zn] = $-$(0.63$\pm$0.16)) for the full system. Using the apparent column density profiles of Si~{\sc ii}, Fe~{\sc ii} and Mg~{\sc i} we show that the depletion and the $N$(Mg~{\sc i})/$N$(Si~{\sc ii}) column density ratio systematically vary across the velocity range. The region with high depletion tends to have slightly larger $N$(Mg~{\sc i})/$N$(Si~{\sc ii}) ratio. The two \hi\ 21-cm absorbers belong to this velocity range. 
The emergence of \zabs = 1.172635 can be understood if there is a large optical depth gradient over a length scale of $\sim$0.35 pc. However, the gas producing the \zabs = 1.173019 component must be nearly uniform over the same scale. 
Systematic uncertainties introduced by the  absorption line variability has to be accounted for in experiments measuring the variations of fundamental constants and cosmic acceleration even when the radio emission is apparently compact as in { PKS~2355-106.}

\end{abstract}




\section{Introduction}
\label{sec_intro}
%

The \hi\ 21-cm absorption is an excellent probe of cold neutral gas. In the Galaxy, observations of \hi\ 21-cm absorption towards high-velocity pulsars and extended sources have revealed structures in the diffuse interstellar medium \citep[ISM; e.g.,][]{Frail94, Heiles97, Brogan05, Roy12}.  These provide important inputs to pressure-equilibrium based models of the ISM.  
Beyond our Galaxy, such studies are limited to a handful of sight lines towards extended radio sources or gravitationally lensed systems or sources with large proper motions \citep[e.g.,][]{Wolfe82, Kanekar01, Srianand13dib, Dutta16, Gupta18j1243}.
The scarcity is due to a combination of small number of known \hi\ 21-cm absorbers and paucity of suitable low-frequency receivers for milliarcsecond scale spectroscopy using the Very Long Baseline Interferometry (VLBI).  It is anticipated that these limitations will be overcome in near future by ongoing Square Kilometre Array (SKA) precursor surveys \citep[][]{Gupta17mals, Allison22} and the availability of SKA-VLBI \citep[][]{Paragi15}.
The variability of extragalactic \hi\ 21-cm absorption lines over the timescales of decades is also relevant for the measurements of variations in fundamental constants of physics and cosmic acceleration \citep[][]{Darling12}.

The MeerKAT Absorption Line Survey (MALS) is one of the large survey projects being carried out with MeerKAT \citep[see][for details]{Gupta17mals}. The survey is well underway and details of the MALS targets and the full survey footprint to search for \hi\ 21-cm and OH 18-cm absorption are summarized in \citet{Gupta21salt}.
As part of MALS science verification, we observed the quasar 
{ PKS~2355-106} (hereafter, \J23). 
This sightline is known to host a damped \lya\ absorber (DLA; $N$(\hi)$>2\times10^{20}$\,cm$^{-2}$) at \zabs = 1.1730  that was pre-selected based on the presence of the strong Mg~{\sc ii} absorption in the SDSS spectrum.
The \hi\ 21-cm absorption  from this absorber was first reported by \citet{Gupta07} using the Giant Metrewave Radio Telescope (GMRT) observations. Subsequently, \citet{Rahmani12} presented a higher resolution GMRT spectrum (0.9 \kms) and an optical echelle spectrum ($\sim$6 \kms) of \J23\ obtained using the Ultra-Violet Echelle spectrograph at the Very Large Telescope (VLT/UVES). They showed that the absorption profiles of singly ionized species like Si~{\sc ii} and Fe~{\sc ii}  span over $\sim$ 150 \kms. They fitted these profiles with 11 independent Voigt profile components with one of them [at \zabs = 1.1730227(29)] coinciding well (i.e within 0.5 \kms) with the redshift of the \hi\ 21-cm absorption [at \zabs = 1.1730188(24)]. This was used to place a stringent constraint on the time variations of $x$ = $g_p \alpha^2/\mu$, where $g_p$, $\alpha$ and $\mu$ are proton g-factor, electromagnetic fine-structure constant and proton-to-electron mass ratio, respectively. None of the other absorption components have shown detectable \hi\ 21-cm absorption in any of the GMRT spectra.


\citet{Ellison2012} observed this quasar with HST/COS (Hubble Space Telescope's Cosmic Origins spectrograph) and confirmed the absorber to be a DLA with log~$N$(HI) = 21.0$\pm$0.1. Assuming radio and optical sightlines are co-spatial, this was used together with the observed \hi\ 21-cm optical depth to constrain the harmonic mean spin-temperature of the gas to \ts$\sim2145\pm570$ K . This suggests that most of the low ionization absorption components seen in the optical spectrum originate from warmer gas or contains only a small fraction of the total $N$(\hi).

The quasar \J23\ is classified as a blazar of flat spectrum radio quasar (FSRQ) type  with detectable $\gamma$-ray emission \citep[][]{Ackermann2015}. The observed synchrotron peak frequency ($\log \nu_p = 12.651$) is consistent with the object being a low frequency synchrotron peaker. While the quasar is classified as non-variable in $\gamma$-rays, it shows substantial variability at 15 GHz \citep[][]{Richards2014} and optical bands \citet[][]{Hovatta2014}. All these are consistent with the presence of a relativistic jet close to our line of sight.

\J23\ has been observed several times with the Very Long Baseline Array (VLBA) of the NRAO\footnote{The National Radio Astronomy Observatory is a facility of the National Science Foundation operated under cooperative agreement by Associated Universities, Inc.} as part of the International Celestial Reference Frame (ICRF) observations in S- (2.3 GHz) and X-bands (8.4 GHz). Based on positional stability and lack of extensive intrinsic source structure, \J23\ is identified as one of the ``defining sources'' of the ICRF catalog \citep{Fey2015}. Gaussian fits recover more than 99\% of the flux density from the central dominant component \citep{Fomalont2000, Fey00}. \citet{Truebenbach2017} have measured a proper motion of $-7.88\pm3.27$ $\mu$as yr$^{-1}$ along the right ascension and $-0.49\pm3.40$ $\mu$as yr$^{-1}$ along the declination.  
%
The VLBA observations at 1.4 GHz \citep{Gupta12} and 605 MHz, i.e., the redshifted frequency of the \hi\ 21-cm absorption \citep{Ellison2012} revealed \J23\ to be compact with almost all the emission being associated with 
an unresolved component. {VLBI space observatory program (VSOP) data of \J23\ obtained at 5 GHz suggests a compact core component of 0.2 Jy within 0.2 mas\citep{Dodson2008}. 
 Recently, \citet{Petrov2017} have reported an off-set of -0.89 mas along RA and -0.40 mas along DEC between VLBA and Gaia coordinates with a probability of 0.29 for this offset to occur by chance. This suggests that the optical and radio sightlines towards \J23\ are most likely co-spatial.}

Here, we present detailed comparison of \hi\ 21-cm absorption profiles obtained towards \J23\ over a period of about 15 years.  
Our high signal-to-noise ratio (S/N) MeerKAT science verification data presented here reveal a new narrow \hi\ 21-cm absorption line within the velocity range covered by the low ionization absorption lines.  We discuss the implications of this to the derived average spin temperature of the absorption system and the nature of components showing \hi\ 21-cm absorption.
Here and throughout the paper we use flat $\Lambda$ cold dark matter ($\Lambda$CDM) cosmology with $\Omega_\Lambda$ = 0.69, $\Omega_m$ = 0.31 and $h$ = 0.674.
All relative velocities used in this work are defined with respect to \zabs = 1.173019. We use solar photospheric abundances from \citet{Asplund2021}.

\section{Observations and Data Reduction}
\label{sec_obs}
%

The field centered at \J23\ was observed using the MeerKAT-64 array as part of MALS science verification on June 02, 2020. The 32K mode of the SKA Reconfigurable Application Board (SKARAB) correlator was configured to split the total observable bandwidth of 544\,MHz centered at 815.9917\,MHz into 32768 frequency channels.  In the vicinity of the redshifted \hi\ 21-cm line frequency corresponding to \zabs=1.173, the channel separation of  16.602\,kHz corresponds to a velocity spacing of 7.6\,\kms.  Of the 64 antennas, 54 participated in these observations.   The total on-source time of 120\,min on \J23\ was split into six scans of 20\,mins at different hour angles to improve the uv-coverage. The total duration of the observing run, which also included two other quasars with known absorption lines, was $\sim$320 min.  The correlator dump time was 8\,sec and the data were acquired for all four polarization products.  J1939-6342 was observed for flux density scale, delay and bandpass calibrations.  The compact radio sources J2348-1631 and J0022+0014 were periodically observed for complex gain calibration.

The MeerKAT data were processed using the Automated Radio Telescope Imaging Pipeline  \citep[ARTIP; we refer the reader to][for details]{Gupta21}.  The steps specific to UHF-band processing of MALS data are described in \citet[][]{Combes21}. The spectral line processing through ARTIP partitions the frequency band into 15 spectral windows (referred as SPW0 to SPW14) with an overlap of 256 frequency channels. Here, we are concerned only with the data products i.e., the Stokes-$I$ continuum image and the continuum subtracted spectral line image cube corresponding to SPW2 covering 629.0 - 667.2\,MHz. The continuum image and the spectral line cube were made using the Briggs' {\tt robust = 0} weighting of the visibilities using the CASA 
task {\tt tclean}.  The continuum image at 648.0\,MHz has a restoring beam of 20.1$^{\prime\prime}\times$17.2$^{\prime\prime}$ with a position angle of $-70.7^\circ$.
%
The rms noise close to the target source is 100\,$\mu$Jy\,beam$^{-1}$. The radio source is compact with a peak flux density of 418.8$\pm$1.0\,mJy, which matches well with the previous estimates of 420-440\,mJy from our GMRT observations.   The quoted uncertainty corresponds to the error from the single Gaussian component fit.  The typical flux density accuracy at these low frequencies is expected to be about $\sim$5\%.  
The rms in the spectrum extracted from the image cube at the pixel corresponding to the peak intensity of the radio source is  0.9\,mJy\,beam\,channel$^{-1}$.

The details of various GMRT spectra used here are provided in our earlier papers \citep[][]{Gupta07, Gupta09, Gupta12, Rahmani12}.
The details of the optical high resolution spectrum of \J23\ used in this study can be found in \citet{Rahmani12}. While \citet{Rahmani12} mainly used these data for constraining variations in fundamental constants of physics, here we use the spectrum to probe the physical state of the gas and inhomogeneity in metal depletion and ionization state.

%
\section{\hi\ 21-cm absorption}
\label{sec_21cm}

\begin{figure}
    \centering
    \includegraphics[viewport=20 155 600 690,width=0.5\textwidth,clip=true]{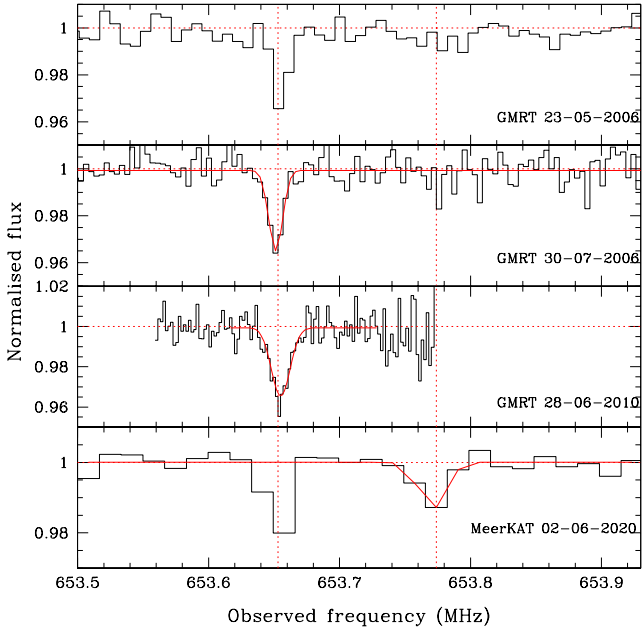}
    \caption{\hi\ 21-cm absorption from the \zabs = 1.1730 DLA towards \J23\ obtained at different epochs, all sharing absorption component-1 at 653.656 GHz. 
    The latest spectrum obtained with MeerKAT shows the additional absorption component-2 at 653.771~MHz.
    Gaussian fits discussed in the text are shown with red curves. 
    }
    \label{fig:21cm}
\end{figure}

As expected, we detect the known \hi\ 21-cm narrow absorption at \zabs = 1.173019 (hereafter ``component-1''). In addition, we also detect another \hi\ 21-cm absorption at \zabs = 1.172635 (hereafter ``component-2''). The normalised spectrum obtained with MeerKAT is shown in the bottom panel of Fig.~\ref{fig:21cm}. This figure also shows the earlier spectra of \J23\ obtained by us with GMRT.  
We also examined the stokes-$I$ spectra of other two quasars included in the same MeerKAT observing run.  These do not show any absorption at 653.771\,MHz whereas the absorption feature consistently appears in the XX and YY spectra of \J23.  Thus, we rule out the possibility of component-2 to be due to Radio Frequency Interference (RFI). 

First we focus on the newly detected \hi\ 21-cm absorption component (i.e., component-2). We fit the \hi\ 21-cm absorption from this component 
with a single Gaussian and obtained \zabs = 1.172635 and FWHM of 11.13 \kms. The integrated optical depth (over the frequency range 653.75-653.80 MHz) is 0.161$\pm$0.030 \kms (i.e., 5.4$\sigma$ detection).  The measured Gaussian FWHM is consistent with the kinetic temperature (and hence the spin temperature, \ts) of the gas being less than 2700\,K.
If we assume the covering factor ($f_c$) of the absorbing gas to be unity (but see below), as
has been used in all previous studies,
we constrain the \hi\ column density in component-2 to $N$(\hi) = 3$\times10^{17} \times$ \ts(K)\  cm$^{-2}$. If we use the upper limit on \ts\ mentioned above we derive \nhi$\le8\times10^{20}$ \cms. 

The frequency range of component-2 is covered in the first two GMRT observations, but not in the highest resolution spectra obtained in our 2010 observations. In the spectrum obtained during 30-7-2006 (second panel in Fig.~\ref{fig:21cm}), that has the best S/N and resolution compared to the one obtained during 23-05-2006, we find the integrated \hi\ 21-cm optical depth of 0.051$\pm$0.039 \kms\ over the above mentioned frequency range for component-2. In this spectrum, the component-2 seen in the MeerKAT spectrum should have been detected at  $>4\sigma$ level.

\begin{figure}
    \centering
    \includegraphics[viewport=20 155 600 690,width=0.5\textwidth,clip=true]{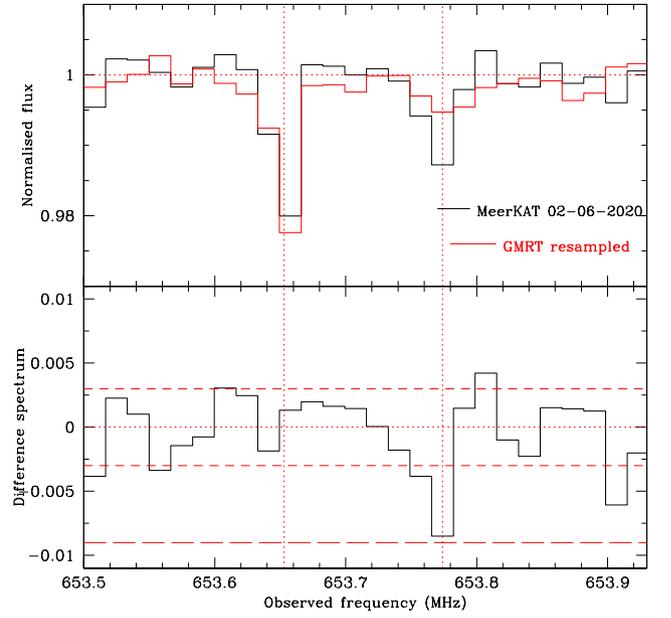}
    \caption{ Comparison of MeerKAT spectrum with combined GMRT spectrum. The latter has been re-sampled to match MeerKAT frequency scale. Bottom panel shows the difference spectrum and the top panel shows the normalized spectra. Vertical dashed lines mark the locations of two \hi\ 21-cm absorption components discussed here. Horizontal short dashed lines in the bottom panel give $1\sigma$ ranges. The long-dashed line shows the 3$\sigma$ level. It is evident that there is a 2.9$\sigma$ level excess detected at the location of peak optical depth for component-2 in our MeerKAT spectrum. 
    }
    \label{fig:21cma}
\end{figure}

To investigate any possible variability in the \hi\ 21-cm optical depth  of component-2, we re-sampled the GMRT spectra (first two epochs) to the frequency scale of the MeerKAT spectrum (shown in Fig.~\ref{fig:21cma}). In the top panel, we compare the two spectra and show the difference between them in the {bottom} panel. It is evident that in the case of component-1 the line intensity variations are well in the $\pm1\sigma$ range (dashed horizontal lines). However, at the location of the absorption peak for component-2 we find the difference 
to be 0.0085$\pm$0.0029 \kms\ (2.9$\sigma$ level).
Therefore, with the present data we confirm the variability at 2.9$\sigma$ level.
Such variability in \hi\ 21-cm absorption is seen towards Blazars \citep[see for example,][]{Briggs83,Kanekar01,Kanekar14} or gravitationally lensed systems \citep[see for example,][]{Muller08, Schulz15} and is primarily related to the small size of the absorbing gas coupled with large proper motion in the background radio source.
We return to this in section~\ref{sec:discussions}.

\begin{table}
    \centering
    \caption{Summary of \hi\ 21-cm absorption optical depth for component-1 
    }
    \begin{tabular}{ccccc}
    \hline
    Telescope & Date & $\delta$v$^a$&${\rm f_{peak}^b}$ & $\int \tau dv^d$\\
              & d-m-y & (\kms)  & (mJy b$^{-1}$)& (\kms)\\
    \hline
    \hline
    GMRT & 23-05-2006   & 3.6 &443 & $0.262\pm0.040$\\
         & 30-07-2006   & 1.8 &420 & $0.240\pm0.037$\\
         & 28-06-2010   & 0.9 & 449$^c$ &$0.295\pm0.022^c$ \\
    MeerKAT &02-06-2020 & 7.9 &419 & $0.221\pm0.019$ \\
         \hline
    \end{tabular}
    \label{tab:tau}
    \flushleft{$^a$ Channel width; $^b$ Peak continuum flux density with an accuracy of about 5\%. $^c$: The data acquired on 28-06-2010 and 29-06-2010 were combined to obtain the continuum image and the spectrum. { $^d$ total optical depth provided is obtained by the direct integration of the optical depth profile.}}
\end{table}

In the following we discuss the absorption from component-1. The \hi\ 21-cm absorption is known to be narrow and sampled only by two channels (each of width 7.6 \kms) in our MeerKAT spectrum.
The details of \hi\ 21-cm absorption at different epochs are summarized in Table.~\ref{tab:tau}. It is also evident from the table that the peak continuum flux density of \J23\ close to the redshifted \hi\ 21-cm frequency has not changed by more than 7\% over the period 2006 - 2020. We do not find any statistically significant optical depth variation between the first two GMRT spectra and our MeerKAT spectrum (typically over an elapsed time of 5.3 yrs in the quasar's frame). 
The table also suggests a $\sim 2.5\sigma$ difference in the \hi\ 21-cm optical depth between our GMRT observations in 2010 and MeerKAT data. However, caution is required since this is likely an artifact of the under-sampling of absorption profile in the MeerKAT spectrum.
A single component Gaussian fit to the highest resolution GMRT data obtained in 2010 yields the line width of $\sigma \sim3.2$ \kms. This provides an upper limit of 1260 K for the kinetic temperature as well as \ts. The maximum $N$(\hi) that can be accommodated in this component is 3.9$\times10^{20}$ \cms. 

The wide frequency coverage of our MeerKAT spectrum allows us to search for \hi\ 21-cm absorption over the redshift range 0.32-1.53. We do not find any other \hi\ 21-cm absorption apart from the above discussed system. In this redshift range we also detect intervening \civ\ absorption systems at \zabs = 1.1624, 1.4101, 1.4106, 1.4111 and 1.4126 in our UVES spectrum. 
Unfortunately our MeerKAT spectrum does not cover the frequency range of {the} associated \hi\ 21-cm absorption with \J23. The presence of associated ionized gas is confirmed by the strong \civ\ absorption seen at \zabs = 1.1624. 
The examination of other spectral windows, reveal non-detections of OH absorption at the systemic redshift of the quasar (or at the redshift of associated \civ\ absorber) and corresponding to the \hi\ 21-cm absorption discussed above.

\section{Associated UV absorption lines}
\label{sec_uvabs}

\subsection{\hi\ absorption and \ts}

As mentioned before, a total \hi\ column density of log~\nhi\ = 21.0$\pm$0.1 was measured using the DLA seen in the HST-COS spectrum. Had we observed \J23\ only with MeerKAT we would have obtained the total \hi\ 21-cm integrated optical depth of 0.384$\pm$0.035 \kms\ 
over the two components.
This gives a \ts$\sim 1419$\,K, which is a factor 1.5 times less than what has been reported by \citet[][]{Ellison2012} (i.e., \ts = $2145\pm570$ K). 
The detection of component-2  suggests the radio and optical sightlines are probing slightly different volumes of gas even when the background source appears to be compact at $\sim$10~mas scales. In that case one can not unambiguously constrain \ts\ even 
when we have total \nhi\ measurements from DLA observations.

\subsection{Metallicity \& depletion}
\begin{figure}
    \centering
    \includegraphics[viewport=30 165 600 690,width=0.5\textwidth,clip=true]{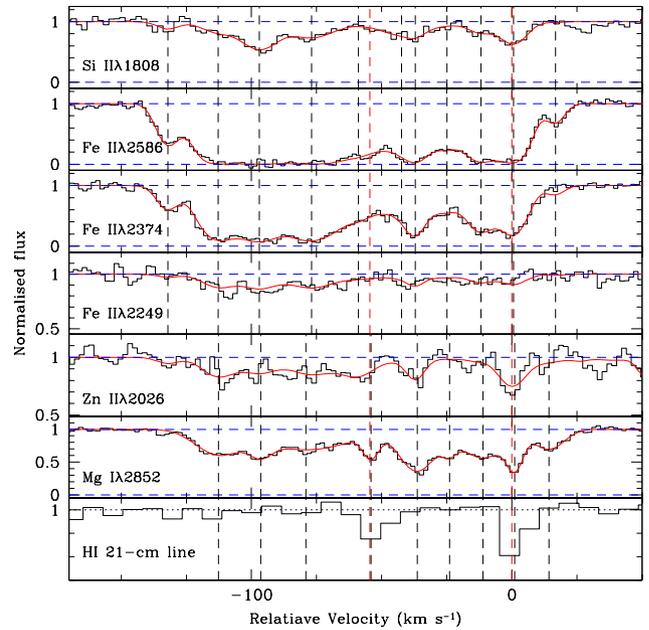}
    \caption{Velocity profile of absorption from different transitions of Si~{\sc ii}, Fe~{\sc ii} and Mg~{\sc i}. The bottom panel shows the velocity plot of \hi\ 21-cm absorption. Black dashed lines identify individual metal line components. 
    Red dashed lines show the location of \hi\ 21-cm absorption. Zero velocity is defined with respect to \zabs = 
    1.173023. It is interesting to note that only Mg~{\sc i} absorption shows distinct component associated with the new \hi\ 21-cm absorption detected in this work (i.e at $-$54.5 \kms; component-2). 
    }
    \label{fig:fe2fit}
\end{figure}

\begin{figure}
    \centering
    \includegraphics[viewport=10 165 600 690,width=0.5\textwidth,clip=true]{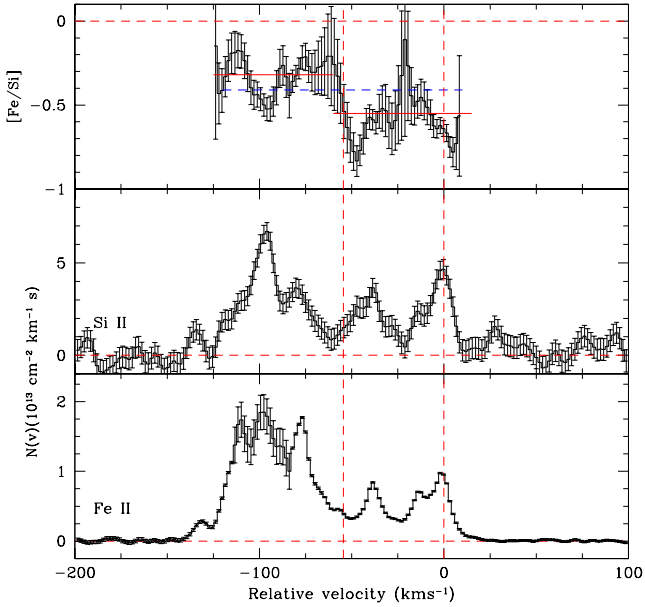}
    \caption{Middle and bottom panel: Apparent column density profile of Fe~{\sc ii} and Si~{\sc ii}. Vertical dashed lines show the velocities at which \hi\ 21-cm absorption is detected. Top panel: Velocity profile of depletion of Fe with respect to Si. It is evident that the depletion is not uniform across the profile. The blue horizontal dotted line gives [Si/Fe] measured based on the total column densities. Red horizontal line gives the same over two velocity ranges discussed in the text. 
    }
    \label{fig:aod}
\end{figure}
If the optical and radio sightlines are different, in a simplistic case, one expects a lack of absorption  at the redshift of component-2 in the optical spectrum. In Fig.~\ref{fig:fe2fit}, we show the velocity plot of \hi\ 21-cm absorption and some metal transitions. Joint fit to singly ionized species performed by \citet{Rahmani12} (red curves in Fig.~\ref{fig:fe2fit}) does not reveal a distinct component at the expected position of component-2. However, we detect a distinct absorption at the expected position in the case of Mg~{\sc i} absorption.  
While decomposing the absorption lines into Voigt profile components is a standard practice, in reality the absorption at a given velocity can be contributed by gas distributed at different physical locations having appropriate radial velocities in projection \citep[see for example,][]{Marra2021}.
Before discussing the origin of component-2 in detail we summarize the basic properties of the
overall system and the two \hi\ 21-cm components obtained using our UVES spectrum.

\citet{Rahmani12} have obtained total column densities using multiple component Voigt profile fitting. The overall profile of absorption lines from singly ionized species are best fitted with 11 Voigt profile components (vertical dashed lines in Fig.~\ref{fig:fe2fit}) and the derived total column densities are, log~$N$(Mn~{\sc ii}) = 12.92$\pm$0.04, log~$N$(Zn~{\sc ii}) = 12.79$\pm$0.04, log~$N$(Si~{\sc ii}) = 15.55$\pm$0.04 and log~$N$(Fe~{\sc ii}) = 15.03$\pm$0.04. When we use log~$N$(\hi) = 21.0$\pm$0.1 and solar abundance from \citet{Asplund2021} we get [Mn/H]=$-(1.50\pm0.11)$, [Zn/H]=$-(0.77\pm0.11)$, [Fe/H]=$-(1.40\pm0.11)$ and [Si/H] = $-(0.96\pm0.11)$. The average depletion across the profile can be quantified using [Fe/Zn] = $-(0.63\pm 0.16)$. We also find the average [Fe/Si] = $-(0.44\pm0.16)$. The depletion measured for Fe, Si and Mn are consistent with what is seen for ``warm disk + halo gas'' of the Milky Way \citep[see figure 6 of][]{Savage96}.

In the case of ``component-1" of \hi\ 21-cm absorption, we find that a metal line absorption component (seen in neutral, singly ionized ions and Al~{\sc iii}) aligns well with the 21-cm absorption (see Fig.~\ref{fig:highions}). For this component, the measured column densities are: log~$N$(Mn~{\sc ii})= 12.10$\pm$0.04, log~$N$(Fe~{\sc ii}) = 14.03$\pm$0.03, log~$N$(Si~{\sc ii}) = 14.75$\pm$0.04,
and
log~$N$(Zn~{\sc ii}) = 12.08$\pm$0.05.
This gives [Fe/Zn] = $-(0.95\pm0.06)$, [Mn/Zn]= $-(0.84\pm0.06)$ and [Fe/Si] = $-(0.64\pm0.06)$.  Thus the depletion in ``component-1'' is higher than the average depletion of the system by $\sim$0.20-0.30 dex. Due to inverse dependence on \ts\,,  the \hi\, 21-cm absorption is biased towards low temperature regions. It is known that \h2\ bearing gas i.e., absorption components having higher density and lower temperature tend to show more depletion compared to the overall metal absorption in DLAs \citep[see for example,][]{Petitjean02,Rodriguez06, Noterdaeme17}.
Thus, our findings are in line with higher depletion in the cooler regions of the absorbing gas.

Multi-component Voigt profile fits to the singly-ionized lines do not identify an independent component at the location of the ``component-2'' of the \hi\ 21-cm absorption.  This suggests that the column density of metal ions is low there, either because of the low $N$(H~{\sc i})  or because of high depletion. 
To explore this we constructed the apparent column density profiles \citep[][]{Savage1991} for Fe~{\sc ii} and Si~{\sc ii}. In the case of Fe~{\sc ii} we use all the available transitions to construct the apparent column density profile capturing the line saturation accurately. These are shown in Fig.~\ref{fig:aod}. In the case of Si~{\sc ii} we mainly use Si~{\sc ii}$\lambda$1808 line. We measure [Fe/Si], assuming these elements are mostly in the singly ionized state, for each velocity bin and plot them in the top panel of Fig.~\ref{fig:aod}. It is evident that [Fe/Si] varies across the profile. The blue horizontal dashed line in this plot shows the average [Fe/Si] obtained using the measured total column densities. We find most of the gas at relative velocities less than -62 \kms\ [this accounts for 68\% of the total $N$(Fe~{\sc ii})] showing [Fe/Si] value above this mean value. Gas at velocities greater than -62 \kms\ [which includes gas producing the two \hi\ 21-cm absorption] show [Fe/Si] below the overall mean. This suggests the presence of gas with two sets of depletion separated in velocity space. Unfortunately we could not perform such an analysis for Zn~{\sc ii}, Mn~{\sc ii} or Cr~{\sc ii} as these absorption lines are weaker and hence errors in $N$(v) are large.

It is also evident from this figure that [Fe/Si] in the velocity range of component-1 is above the mean value measured over the full profile. The gas in the relative velocity range $>-$62 \kms\ 
is consistent with the mean [Fe/Si] measured (red horizontal line). This is not the case at the location of component-2. However the maximum depletion [Fe/Si] is measured within 6 \kms\ to this component. In summary, within measurement uncertainties, we identify two distinct velocity ranges over which the depletion pattern is well separated.
Both \hi\ 21-cm absorption discussed here are in the velocity range showing higher depletion.

\subsection{Ionization state}

\begin{figure}
    \centering
    \includegraphics[viewport=10 165 600 690,width=0.5\textwidth,clip=true]{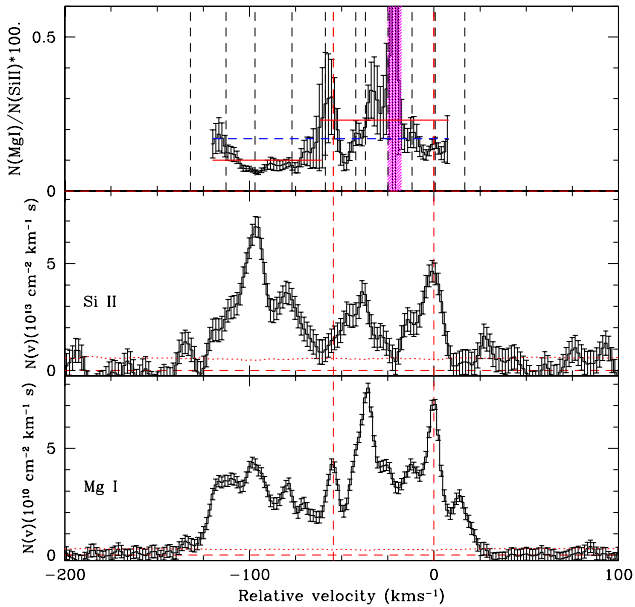}
    \caption{Middle and bottom panel: Apparent column density profile of Si~{\sc ii} and Mg~{\sc i} absorption. Red vertical dashed lines show the velocity at which \hi\ 21-cm absorption is detected.
    Red dotted lines give 1$\sigma$ error. Top panel: The column density ratio $N$(Mg~{\sc i})/$N$(Si~{\sc ii}) as a function of velocity. The error bar gives the 1$\sigma$ error in the ratio. It is evident that this ratio is not uniform across the profile. The blue horizontal dashed line gives median ratio measured based on the total column density and the red horizontal line gives the same over two velocity ranges discussed in the text. 
    }
    \label{fig:mg1bysi2}
\end{figure}

\begin{figure}
    \centering
    \includegraphics[viewport=10 165 600 690,width=0.5\textwidth,clip=true]{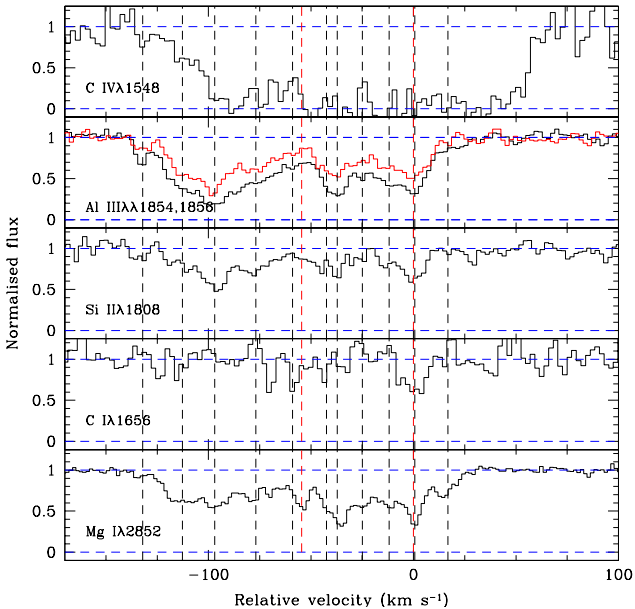}
    \caption{Velocity plot showing the profiles of different ionized species. Ticks are as in Fig.~\ref{fig:fe2fit}}.
    \label{fig:highions}
\end{figure}

The ratio of Mg~{\sc ii} to Mg~{\sc i} column densities is governed by the electron density, $n_e$, gas temperature and photoionization rate, $\Gamma_\nu$. 
As Mg~{\sc ii} lines are highly saturated one can use Si~{\sc ii} column density for this purpose.
In Fig.~\ref{fig:mg1bysi2} we plot the apparent column density profile of \mgi\ and Si~{\sc ii} and their ratio (over the velocity range where Si~{\sc ii}$\lambda$1808 line is detected). Like in the case of depletion, this ratio shows a clear trend. In the velocity range $<$-62 \kms\ the ratio $N$(Mg~{\sc i})/$N$(Si~{\sc ii}) tends to be lower than the mean value. Whereas the ratio tends to be higher than the mean value for higher velocities (i.e $>$-62 \kms).

In the second panel from the bottom of Fig.~\ref{fig:fe2fit} we show the fits to the Mg~{\sc i} absorption lines.
The Voigt profile fit to Mg~{\sc i} was performed independent of the fits to the singly ionized species. Interestingly, Mg~{\sc i} absorption profile follows most of the strong components of the singly ionized species. 
The total column density of Mg~{\sc i} is log~$N$(Mg~{\sc i}) = 12.85$\pm$0.07. This gives the average 
log~[$N$(Mg~{\sc i})/$N$(Si~{\sc ii})] of
$-(2.70\pm0.13)$. 

From the bottom panel it is evident that Mg~{\sc i} absorption is detected for ``component-1'' of \hi\ 21-cm absorption. We measure, log~$N$(Mg~{\sc i}) = 12.35$\pm$0.13.
This gives log~[$N$(Mg~{\sc i})/$N$(Si~{\sc ii})]=$-2.40\pm 0.13$ for this component. This means that the gas responsible for ``component-1'' on average has at least a factor 2 lower value (albeit with large errors) of $\Gamma_\nu/(n_e \alpha)$ with $\alpha$ being the recombination rate.
Clear detection of C~{\sc i} absorption (see Fig.~\ref{fig:highions}), with log~$N$(C~{\sc i}) = 12.96$\pm$0.05, only from this component supports this conclusion. 
We also obtain a 3$\sigma$ upper limit of log~$N$(C~{\sc i*})$\le$12.44. Note the upper limit on the column density ratio $N$(C~{\sc i*})/$N$(C~{\sc i}) is consistent with what has been measured for H$_2$ bearing DLA components at high-z \citep[see figure 12 of][]{Srianand05}. We obtain an upper limit of hydrogen density ($\mathrm{n_H}$ $\le$ 51 cm$^{-3}$)  by assuming a gas temperature of 100~K, the excitation is mainly due to cosmic microwave background and collisions with neutral hydrogen \citep[cross sections taken from][]{Launay1977} and the spontanous decay is the  only deexcitation process. 

For component-2 we measure log~$N$(Mg~{\sc i}) = 11.59$\pm$0.23 and log~$N$(C~{\sc i})$\le$12.34 ($3\sigma$). The large error in $N$(Mg~{\sc i}) comes from the large error in the b-parameter due to the line being narrow and blended with other components. From the Si~{\sc ii}$\lambda$1808 line profile we obtain an upper limit, log~$N$(Si~{\sc ii})$\le$14.25. This suggests log [$N$(Mg~{\sc i})/$N$(Si~{\sc ii})]$\ge-$2.66. This limit allows for the physical conditions in component-2 to be similar to that of component-1. 

It is evident from this figure that the \mgi\ absorption component at the relative velocity of $\sim$14.5 \kms\ does not show detectable Si~{\sc ii}$\lambda$1808 absorption (see also Fig.~\ref{fig:fe2fit}). Based on the voigt profile fits we find  log [$N$(Mg~{\sc i})/$N$(Si~{\sc ii})] = $-2.15\pm$0.28. Despite this we do not detect \hi\ 21-cm absorption from this component. This component contains only $\le$ 1\% of the observed total column density of Si~{\sc ii} and Fe~{\sc ii}. Thus the non-detection of \hi\ 21-cm absorption from this component with a signature of low ionization could be related to low total \hi\ column density.

In Fig.~\ref{fig:highions} we compare the velocity profile of absorption from different ionization states. The absorption profiles of Al~{\sc iii} roughly follow the singly ionized species. 
As typical of DLAs \citep[][]{Fox07b}, the \civ\ profile is much wider and not necessarily following the low ionization species. However, what is more interesting is that the \civ\ profile is symmetric around the low ionization components that show higher depletion and low ionizing conditions (i.e $-$62 to +10 \kms). In the velocity range $-$120 to 100 \kms the \civ\ absorption seems very weak while in the velocity range 20-50 \kms \civ\ absorption is detected without strong absorption from low ionization species.

\section{Results and discussions}
\label{sec:discussions}
 Time variability of
\hi\ 21-cm absorption is a good tool to probe the small scale structures in the absorbing gas. 
The present understanding is that the ISM is
a magnetized medium where the turbulent cascade, driven by a local energy source and acting jointly with phenomena such as thermal instability, is the source of these structures \citep[see][for a detailed review]{Stanimirovic2018}.
Similarly, VLBA spectroscopy has revealed parsec scale structures in some intervening \hi\ 21-cm absorbers \citep[see][]{Srianand13dib}.
Presence of very compact cold \h2\ bearing components in some DLAs is established through partial coverage \citep{Balashev2011, Klimenko2015} and time-variability \citep{Boisse2015} in the optical spectrum.

Here, we report the emergence of a new \hi\ 21-cm absorption component at \zabs =  1.172635 towards the blazar \J23.
There are currently three known cases of variability in the \hi\ 21-cm absorption from  high redshift intervening absorbers. All of them are toward blazars. These are (i) \zabs = 0.525 toward the BL Lac object AO 0235+167 \citep{Wolfe82}, (ii) \zabs = 0.3127 toward FSRQ PKS 1127-145 \citep{Kanekar01} and (ii) \zabs = 2.0395 toward FSRQ PKS 0458-020  \citep[][]{Kanekar14}.
There are {a few} things in common between these DLAs: (i) large measured $N$(\hi) compared to typical DLAs (i.e log~$N$(\hi)$\sim$21.70); (ii) metal lines are wide spread (with $88\le \Delta V_{90}\le 120$ \kms) and (iii) the background sources are detected in $\gamma-$ray emission.

In the case of AO 0235+167 and PKS 1127-145, the \hi\ 21-cm absorption is detected in several components spread over $\sim$ 100 \kms. All these components show time-variability in the \hi\ 21-cm  optical depth without significant velocity shifts over a range of time-scales. In the case of PKS 0458-020, the \hi\ 21-cm absorption is spread over $\sim$50 \kms\ and the quoted variability is based on three epoch observations. 
Multiple component structures at the mas scale are clearly visible in the case of FSRQs PKS 0458-020 and PKS 1127-145 \citep[see for example,][]{Fomalont2000}.
The absorber towards \J23\ also exhibits widespread metal absorption lines and high $N$(H~{\sc i}).  The quasar is also detected at $\gamma$-rays.  The only difference with respect to above mentioned three cases is the narrow width i.e.,  $\Delta V_{90}$ = 12.5\,\kms\ (component-1) of \hi\ 21-cm absorption.

There are three possibilities considered in the literature to understand variability in \hi\ 21-cm absorption. These are (i) changes in the line-of sight column density due to the proper motion of the background source; (ii) inter-stellar scintillation (ISS) and (iii) 
changes in the physical conditions in the absorbing gas. Given the time-scales involved and the low transition probabilities of \hi\ 21-cm line driving the changes in \ts\ i.e., the possibility (iii) is unlikely. As the radio continuum at the redshifted 21-cm wavelength has not varied by more than 7\% during our observations we conclude that the observed variability is not driven by ISS.

Here, we consider only the possibility-(i) i.e., optical depth variations caused by the superluminal motion of radio components associated with the background radio source i.e., blazar.  This is motivated by the core-jet morphology exhibited by \J23\ at high frequencies.  In the VLBA images at 5\,GHz and 8.6\,GHz, 98\% of the total flux density is contained the radio core and there is an indication of a weak component 1-2\,mas (8-15\,pc) away \citep[][]{Fomalont2000}.  An additional component 8\,mas from the core is also detected in lower frequency 2.3\,GHz image \citep[][]{Fey00}.  The radio structure is unresolved in the 1.4\,GHz VLBA image with a deconvolved size of $<$30\,pc \citep[][]{Gupta12}.  The variations in substructure in radio continuum at parsec scales, which may be more prominent at lower frequencies due to the steeper spectral index of the jet component, may be responsible for observed variation in \hi\ 21-cm absorption profile.

Indeed,  \citet{Truebenbach2017} measured a proper motion of $\Delta$RA = $-7.88\pm3.27$\,$\mu$arcsec yr$^{-1}$ and $\Delta$DEC = $-0.49\pm3.40$ $\mu$arcsec yr$^{-1}$.
The time-elapsed in the quasar frame between our GMRT and MeerKAT observations is $\sim 5.3$ yrs. We expect a displacement of 
$\sim$ 40 $\mu$arcsec during this period. 
Such a motion corresponds to a shift in the line of sight of 0.35 pc at the redshift of the \hi\ 21-cm absorber, implying a strong gradient in \hi\ 21-cm optical depth for  component-2 over this scale.
Meanwhile, the absorbing gas correspoding to component-1 would have no such large gradient in optical depth over similar scales. Uncorrelated variability between different \hi\ 21-cm absorption components have been seen in previous cases, suggesting the presence of structures at different spatial scales in the absorbing gas. 

\begin{figure}
    \centering
    \includegraphics[viewport= 25 140 600 700,width=0.5\textwidth,clip=true]{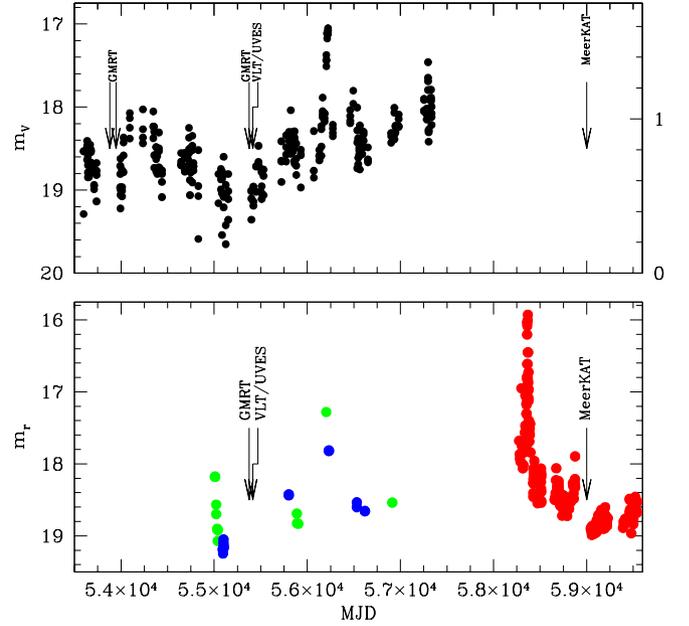}
    \caption{ {\it Top panel:} CRTS V-band light curve together with S (2.3 GHz; blue) and X (8.6 GHz, magenta) band VLBA light curves. {\it Bottom panel} The r-band light curve of \J23\ based on available data from ZTF (red points), PTF (green points) and PanSTARRS(blue points). Arrows mark the epoch of spectroscopic observations during the same period.}
    \label{fig:ztflc}
\end{figure}

The publicly available optical light curves from Zwicky transient factory \citep[ZTF;][]{zptf2019a,zptf2019b} also show that \J23\ has gone through an eruption event around July 2018 when the quasar got brighter by about 2 magnitudes in both $g$- and $r$-bands (shown in Fig.~\ref{fig:ztflc}). While our MeerKAT spectroscopic observations occurred a couple of years after that event, all the VLBA low frequency continuum observations that we use to discuss the radio morphology were obtained much before that event. Therefore, it will be interesting to see whether there was any structural change in the radio emission of \J23\ after the strong brightening seen in July 2018. 

If synchrotron emission has appreciable contributions to the optical continuum emission then we expect the UV absorption lines to show time variability.
While the presence of broad emission lines may indicate the UV continuum in \J23\ being dominated by disk emission, it will be interesting to investigate the time variability using  high resolution optical spectrum. In particular it is interesting to investigate the presence (and variability) of partial coverage.

The anticipated increase in the number of \hi\ 21-cm absorbers and the availability of SKA will make it feasible to carry out large scale studies to detect variations in fundamental constants of physics and cosmic acceleration. 
Outcome of these experiments depends on high frequency stability of the instruments and absorption features. Presence of variability in absorption profiles will introduce systematic uncertainties. This can be minimized by avoiding H~{\sc i} 21-cm absorbers detected towards blazars and use large sample of systems. In the latter case one hopes the systematics introduced is random and gets canceled when we have a large sample. The \hi\ 21-cm absorption surveys being carriedout using MeerKAT and ASKAP will provide such a large sample.
\section*{Acknowledgements}

We thank MeerKAT and GMRT staff for their support during the observations.  The  MeerKAT  telescope  is  operated  by  the  South  African  Radio  Astronomy  Observatory,  which  is  a  facility of   the   National   Research   Foundation,   an   agency   of   the   Department   of   Science   and Innovation. GMRT is run by the National Centre for Radio Astrophysics of the Tata Institute of Fundamental Research.  The MeerKAT data were processed using the MALS computing facility at IUCAA (https://mals.iucaa.in/releases). KM acknowledges support from the National Research Foundation of South Africa.

\section*{Data Availability}

The data used in this work are obtained using MeerKAT (SSV-20180516-NG-02)
and GMRT. Raw data will become available for public use in
accordance with the observatory policy. The MeerKAT data
products will be publicly released through the survey website:
https://mals.iucaa.in.



\bibliographystyle{mnras}
\bibliography{J2358_mals} 






\bsp	
\label{lastpage}
\end{document}